# Response to Comment on "Stellar activity masquerading as planets in the habitable zone of the M dwarf Gliese 581"

Paul Robertson[1,2], Suvrath Mahadevan[1,2,3], Michael Endl[4], Arpita Roy[1,2,3]

**Abstract:** Anglada-Escudé and Tuomi question the statistical rigor of our analysis while ignoring the stellar activity aspects that we present. Although we agree that improvements in multiparametric radial velocity (RV) modeling are necessary for the detection of Earth-mass planets, the key physical points we raised were not challenged. We maintain that activity on Gliese 581 induces RV shifts that were interpreted as exoplanets.

Anglada-Escudé and Tuomi (1) raise some interesting and valid points regarding statistical analysis of radial velocity (RV) data. Although we heartily agree on the ultimate statistical desirability of fitting all RV signals simultaneously, their Comment misses the heart of the matter regarding the necessity to first find ways to discriminate stellar activity from bona fide exoplanets—i.e., our findings of stellar activity in Gliese 581 (2). It is important to reemphasize that the primary argument of our analysis was not statistical but physical. Our work clearly established the rotation period of GJ 581 to be 130 days—twice the period of "planet d" independently of the statistical arguments raised in the Comment. Without compelling evidence otherwise, a periodic RV signal at the stellar rotation period or its integer fractions should be assumed to be associated with stellar activity instead of a planet [as demonstrated by (3), among others]. As recently noted by the High Accuracy Radial Velocity Planet Searcher (HARPS) team—who originally discovered "planet d"—had the rotation period of GJ 581 been previously known, it is unlikely that the 66-day RV signal would ever have been ascribed to an exoplanet (4). These authors identify an analog to GJ 581 in the GJ 3543 system and attribute the detection of an RV signal at half the rotation period to stellar activity, in the absence of definitive evidence favoring the exoplanet hypothesis.

Our analysis, although statistically and conceptually simple, strongly disfavors the planetary interpretation for GJ 581d and g. The connection between the stellar magnetic activity revealed in the Hα line and the RV signals of these planets is evident using the same statistical techniques as the Gliese 581 planet discovery papers (5–7), even without an exhaustive statistical exploration of the full parameter space. The field of RV exoplanet detection needs to make simultaneous progress on three fronts in order to detect low mass planets in the habitable zones of the nearest stars:

1) Better RV precision with new instrumentation, calibration, and reduction techniques.
2) A better understanding of stellar activity, activity indicators, and ways to ameliorate its impact. This may be done by treating the problem (e.g., this work) or by avoiding it [e.g., by using near-infrared spectra (8)].
3) Improvements in statistical techniques to not just detect signals but also to be able to disentangle Doppler RV signals from stellar activity–induced signals.

[1] Department of Astronomy and Astrophysics, The Pennsylvania State University, University Park, PA 16802, USA. pmr19@psu.edu

[2] Center for Exoplanets and Habitable Worlds, The Pennsylvania State University, University Park, PA 16802, USA.

[3] The Penn State Astrobiology Research Center, The Pennsylvania State University, University Park, PA 16802, USA.

[4] McDonald Observatory, The University of Texas at Austin, Austin, TX 78712-1206, USA.

We agree that the ideal way to model RV data for multiplanet systems with stellar activity is to simultaneously model Keplerian orbits and activity signals/indicators as a single, multiparametric fit. We are well aware of these issues, as we have ourselves stated in another recent paper (9). There are a number of technical and scientific challenges to such an approach, most notably how to correctly parameterize and perform such a fit and how to best invoke the Gaussian process framework, which has been used successfully to describe stellar activity in recent work (10). The insidious impact of stellar activity—even with relatively inactive stars like Gliese 581—is only now becoming apparent, and no widely recognized or accepted framework currently exists for treating it. As a first approach, our paper relied on the same statistical techniques as used by the studies that resulted in the original discoveries of the GJ 581 planets to show that accounting for activity already explains many of the puzzling aspects of the existing RV data. We and other members of the exoplanet community are eagerly working toward developing and implementing a more complete treatment.